\documentstyle[preprint,aps]{revtex}
\begin{document}
\draft
\title{ Staggered Superconductivity in UPt$_3$:  A New Phenomenological
Approach }
\author{R. Heid, Ya. B. Bazaliy, V. Martisovits, and D. L. Cox}
\address{ Department of Physics, The Ohio State University, Columbus,
Ohio 43210 }
\maketitle
\mediumtext
\begin{abstract}
We present a new Ginzburg-Landau theory for superconductivity in
UPt$_3$, based upon a multicomponent order parameter transforming under
an irreducible space group representation; the phase is staggered in
real space.  Our model can explain the $H-T-P$ phase diagram including
the tetracritical point for all field directions.  We motivate this
unconventional superconducting state in terms of odd-in-time-reversal
pairing that may arise in one- or two-channel Kondo models, and suggest
experimental tests.
\end{abstract}
\bigskip
\bigskip
\narrowtext

%-------------- begin text ------------------------
Heavy fermion superconductors have continued to attract attention since
their discovery
in 1979\cite{hfrevs}.  In these materials, fermionic excitations with
effective masses hundreds of times the free electron mass undergo a
pairing transition.  Anomalous power laws in low temperature
thermodynamic and transport properties\cite{hfrevs} together with
complex
phase diagrams suggest interpretation in terms of Cooper pairs with
exotic
symmetries.
For example in the magnetic field-temperature $H-T$ plane for
ambient pressure $P\approx 1$ bar, the hexagonal material
UPt$_3$ apparently possesses three distinct superconducting phases,
shown as $A,B,C$ in Fig.~1\cite{adenwella89}.
In the $P-T$ plane for
zero $H$ there are similarly three observed phases\cite{upt3press}.
Intriguingly, the $B$ phase in
UPt$_3$
appears to break time reversal symmetry ${\cal T}$ in that
muon spin rotation spectroscopy reveals the development of tiny
additional magnetic moments (order $0.001\mu_B$)\cite{upt3musr}.
So far, no comprehensive microscopic theory exists for these
materials and so much theoretical effort has been devoted to
phenomenological approaches.  No such approach to date has 
successfully explained UPt$_3$ without substantial fine tuning
of parameters.  

In this Letter, we propose a new model for the phase diagram of UPt$_3$
in which finite center of mass momentum (FCM) pairs give rise to an
order parameter which transforms as a triplet under the operations of
the full hexagonal space group.  
Our model produces a
plausible description of the $H-P-T$ superconducting phase diagram. 
In particular, we can explain the tetracritical point found  
for all field orientations.  While the Ginzburg-Landau free
energy contains numerous parameters, experiment constrains the zero
field values to a narrow region which allows for a
${\cal T}$-breaking $B$ phase. The crystalline structure of UPt$_3$
plays a crucial role in our considerations.  
The proposed superconducting order parameters are
microscopically motivated in terms of odd-in-${\cal T}$
pairing\cite{baletal} such as may arise in the two-channel
Kondo model\cite{emkiv,affludgrn2} suggested for heavy fermion
materials\cite{coxtwoch}.  Such pairs may
generically give rise to negative pair hopping energy favoring
staggered superconducting states\cite{mirtsvcole}.  We shall discuss
data supporting this odd-in-${\cal T}$ pairing and point out crucial
tests of our model as well as some of its limitations.
We note that our theory is applicable only near $T_c$ and $H_{c2}$, 
and is macroscopic in
character.  Explanation of the low temperature power laws awaits 
the development of a suitable microscopic approach. 

To place our work in context, we first discuss the $E$ doublet
model\cite{saulsrev}.
In this picture, the Cooper pairs have zero center of mass momentum
(ZCM), and are described by basis functions $\phi_a(\vec
k),\phi_b(\vec k)$ which form a closed vector space under the 
symmetry operations of the hexagonal space group.  With each
basis function is associated an order parameter amplitude
$\eta_{a,b}$, defined through 
the gap function 
$\Delta(\vec k)\sim \eta_{a}\phi_a(\vec k)+\eta_{b}\phi_b(\vec k)$.
The phenomenological approach consists of writing down and 
minimizing a Ginzburg-Landau
free energy which is an expansion in the $\eta$ amplitudes.   A zero
field splitting is present which most likely originates in the
orthorhombic strain induced by magnetic order. Then the three
phases have
 $\eta_a\ne 0$, $\eta_b=0$($A$-phase), $\eta_a\ne 0$, $\eta_b\ne
 0$($B$-phase),
$\eta_a=0$, $\eta_b\ne 0$($C$-phase). 
The apparent time reversal breaking
in the $B$-phase is explained by suitable a choice of Ginzburg-Landau
parameters which energetically favor a relative phase factor of $i$
between $\eta_a,\eta_b$.
The $E$ model has a flaw: while the data show a tetracritical point for
all field orientations,  a
tetracritical point generically arises
only for in-plane magnetic fields\cite{chengarg}.
This problem is remedied by fine tuning to zero the gradient
term in the free energy which mixes the $a,b$
components\cite{saulsrev}.
As a possible alternative, the authors of Ref.
\cite{chengarg} propose a model of accidentally nearly degenerate
ZCM 
singlet states (DS model) which have no such gradient mixing term. 
While this can 
explain the tetracritical
point for all field orientations, there is no compelling
explanation for the accidental near degeneracy.

Our work proceeds via a different phenomenological route:  by
considering Cooper pairs with FCM, we are free to examine space group
representations which merge the desirable properties of the $E$
doublet and DS models by having order parameter 
degeneracy and the absence of
gradient mixing terms enforced by symmetry. 

Before turning to the core of the paper, we shall briefly review the
motivation for this work in terms of odd-in-${\cal T}$ pairing theory.
For an odd in ${\cal T}$ pair wave function it may be shown within a
quasiparticle framework\cite{baletal,mirtsvcole} that (i) for ZCM
pairs, the ``Meissner
stiffness'' which goes as $1/\lambda_L(T)^2$, $\lambda_L$ the London
penetration depth, is
{\it negative}; (ii) an immediate corollary is that the pair transfer
energy between two odd-in-${\cal T}$ slabs, or Josephson coupling, is
negative--alternatively, the coefficients of the gradient terms in the
free energy are negative.  To the extent that these properties are
generically true (it is unknown whether a quasiparticle picture is
applicable in all cases), we then expect a non-uniform superconducting
state in which the pairs have FCM and the phase is modulated in real
space.  Indeed, the Majorana Fermion treatment of the ordinary Kondo
lattice model favors just such a state, with the superconducting order
parameter fully staggered, i.e., the pair hopping from site to site is
negative\cite{mirtsvcole}.

There is so far no numerical evidence supporting this
odd-in-${\cal T}$ picture for the single channel Kondo 
lattice model\cite{jarrtobe}. 
There are strong reasons to consider this state
in the two-channel Kondo lattice model. In this model, two identical
``channels'' of spin 1/2 conduction electrons interact
antiferromagnetically with a lattice of spin 1/2 local moments.   For
this model it has been shown that  (i) in the impurity limit, the pair
field susceptibility for the odd-in-${\cal T}$ order parameter which is
a singlet in spin and channel indices diverges as $-\ln(T),~~T\to 0$,
and (ii) all heavy fermion superconductors have appropriate symmetry
and dynamical conditions to be describable by two-channel Kondo
models\cite{coxtwoch,coxsend}.  For the uranium based materials a
two-channel quadrupolar Kondo model can apply, with spin corresponding
to local quadrupolar indices and the channel labels corresponding to
conduction magnetic indices.  
The spin and channel
singlet states transform as the  $A_2$ representation for each point
group relevant to the heavy fermion materials\cite{coxludtobe} ($A_2
\sim (x^3-3xy^2)(y^3-3yx^2)$ for the hexagonal point group).

Assuming that such local odd-in-${\cal T}$ pairs of $A_2$ point
symmetry form at each site 
with negative pair hopping in the heavy fermion
superconductors, we can determine the allowed multidimensional 
space group representations
which describe the ordering in each compound.  Note that such a
description is reasonable only to the extent that the Kondo screening
length is strongly reduced in the lattice relative to the impurity,
which is in fact expected in the single channel Kondo lattice
model\cite{millee}.  In a hexagonal crystal with one 4f/5f atom per
unit cell the pair energy minima will be at the two $K$
points of Fig.~2.  For UPt$_3$, the crystal structure with two atoms
per unit cell and interlayer frustration for antiferromagnetic (phase)
couplings pushes the pair hopping minima away from the $K$ points, and
for a wide range of parameter values they move to the $M$ points which
have a three point star\cite{also}.   In addition,
the presence of two atoms per unit cell gives rise to non-degenerate
bonding and non-bonding combinations of pair orbitals within each
cell.
We assume that the non-bonding combination is sufficiently high in
energy to be negligible in the current analysis.  

%------------------- model

We now turn to the discussion of a specific Ginzburg-Landau model for
UPt$_3$ based upon the 3 dimensional $M$-point representation.  While
motivated by the odd-in-${\cal T}$ order parameter discussion above, we
are not limited to this as a potential microscopic source for our
phenomenology.
Free energy invariants are constructed by standard group theoretical
methods, and
we have checked the completeness of our set using image group
techniques\cite{landauref}.
We include a linear coupling to an orthorhombic strain field
$\epsilon$,
which breaks the hexagonal symmetry.
The free energy density is given by
\begin{mathletters}
\begin{eqnarray}
F&=&F_0+F_g^{ab}+F_g^c \\
F_0&=&\alpha_+ |\eta_3|^2 +\alpha_- (|\eta_1|^2+|\eta_2|^2)
+\beta_1 \sum_n |\eta_n|^4 \nonumber \\
&+& \beta_2 \sum_{n\neq m} |\eta_n|^2
|\eta_m|^2
+\beta_3 \sum_{n\neq m} \eta_n^2 \eta_m^{*2} \\
F_g^{ab}&=&\sum_n \{\mu_n (|p_+ \eta_n|^2 + |p_- \eta_n|^2)
+ \mu^\prime_n (|p_+ \eta_n|^2 \nonumber \\
&-& |p_- \eta_n|^2)
+(\nu_n (p_+ \eta_n) (p_-\eta_n)^* + c.c.)\}  \\
F_g^c&=&\sum_n \chi_n|p_z \eta_n|^2
\label{model}
\end{eqnarray}
\end{mathletters}
Here, $\eta_n$, n=1,2,3, are the order parameter components,
$p_\pm=(p_x\pm i p_y)/\sqrt{2}$, $p_\alpha=-i\partial_\alpha - 2 e
A_\alpha / \hbar c$ ($e<0$), and $\alpha_\pm=a_0 (T-T_{\pm})$ with
$T_{+}=T_{c0}+2  \epsilon/a_0$ and
$T_{-}=T_{c0}- \epsilon/a_0$.
For the odd-in-${\cal T}$ order parameter considered in the frequency
domain, the $\eta_i$ are understood to be the leading order 
coefficients of the expansion about $\omega = 0$\cite{baletal}.
We will restrict
further analysis to the case $\epsilon > 0$ ($T_+ > T_-$),
which is a necessary condition
to comply with the suggested ${\cal T}$-breaking in phase B
but not in phase A.
The coefficients of the gradient terms are expanded to linear order in
$\epsilon$ and are defined by
a number of phenomenological coupling constants $\kappa_j$ through
$\mu_n=\kappa_1+\kappa_2 \epsilon Re(\gamma^n)$,
$\mu^\prime_n=\kappa_3 \epsilon Im(\gamma^n)$,
$\nu_n=\kappa_4 \gamma^n +\kappa_5 \epsilon+\kappa_6 \epsilon
\gamma^{-n}$, and
$\chi_n=\kappa_7+\kappa_8 \epsilon Re(\gamma^n)$.
The phase factor $\gamma=exp(i 2 \pi /3)$ reflects the three-fold
symmetry of the star of M.
Note that the last term in Eq.\ (1b) favors ${\cal T}$-breaking
solutions for $\beta_3 > 0$.

Model (1) accounts for a rich phase diagram already when $H=0$. 
Fig.~3 summarizes the stability region of zero field solutions.
Four regions in the parameter space can be distinguished according to
their predicted phase sequence, as shown in Table 1 \cite{regionlines}.
Most promising with regard to experiment are regions I and II.
Both allow for two second order transitions at $T_+$ and
$T_*$($<T_+$),
where the second phase manifestly breaks ${\cal T}$-symmetry and can be
identified with phase $B$ of Fig.~1b.

For region I, we recover the experimental $P-T$ diagram,
when we
assume a pressure dependent strain field, which decreases with
increasing $P$ and vanishes for $P>P_c$ \cite{bcline}.
On the other hand, a third low-temperature phase for finite $\epsilon$
is predicted for region II, which signals an unlocking of the relative
phases of the order parameter components.
Consequently, a fourth superconducting phase $D$  appears between 
$B$ and $C$,
and all phase boundaries meet at a pentacritical point.
Although this scenario seems to be at variance with Fig.~1b, we cannot
immediately reject it because the $BD$ and $DC$ boundaries are
predicted to be very close together rendering an observation of the
fourth phase difficult.
Estimates for the model parameters, as derived from experimental data
(shaded area inside the region of stability in Fig.~3)
do not unambiguously favor one of the two regions.

Let us now discuss the $H-T$ diagram predicted by model (1).
Generally, the upper critical field $H_{c2}$ is determined by the
smallest eigenvalue of the linearized Ginzburg-Landau equations
\cite{chengarg,garg}, which can be seen as a coupled system of
Schr\"odinger-like equations, with the order parameter playing the role
of a multicomponent wavefunction.
In our case, this leads to three {\em decoupled} equations, which makes
it possible to obtain all eigenvalues in a closed analytic form for a
general magnetic field.
For ${\bf H}=H(\cos \varphi \sin \theta, \sin \varphi \sin \theta, \cos
\theta)$,
$H_{c2}$ is given by the largest of the three values
\begin{eqnarray}
H_n(T) = \frac{\hbar c}{2 |e|} \frac{-\alpha_n(T)}{\sqrt{R_n}+\mu_n^\prime
\cos \theta}    \nonumber \\
R_n= (\mu_n^2-|\nu_n|^2) \cos^2 \theta +\chi_n(\mu_n-Re(\nu_n
e^{i2\varphi})) \sin^2 \theta
\end{eqnarray}
where $\alpha_3\equiv \alpha_+$ and $\alpha_1=\alpha_2\equiv
\alpha_-$.

A tetracritical point in the $H-T$ plane corresponds to a crossing of
the
two largest values $H_n$ as a function of temperature.
As mentioned by Garg \cite{garg}, such a crossing necessarily requires
the existence of a conserved quantity associated with the Schr\"odiger
equations.
For our model, this property follows immediately from the
decoupling of the equations dictated by translational invariance,
therefore allowing, in principle, a
tetracritical point for {\em all} directions of the magnetic field.
Within a one domain model,
the almost complete isotropy of $H_{c2}$ in the  basal plane
\cite{brison94}
requires that the in-plane anisotropic gradient terms are very small
($|\nu_n|\ll\mu_n$).
For the extreme case $\nu_n=0$, we find kinks in $H_{c2}$ for all field
directions provided that $\kappa_1>0$, $\kappa_7>0$,
$\kappa_2+|\kappa_3|/\sqrt{3}>0$, and
$\kappa_2 \kappa_7+\kappa_1 \kappa_8>0$ \cite{hc2cond}.
These constraints do not represent strong limitations.
Note, that the gradient terms proportional to $\epsilon$ are necessary
for a tetracritical point to appear because for vanishing strain the
slopes of the $H_n$ versus $T$ lines would be identical.
As a consequence, the slopes of the low-field and high-field
$H_{c2}$-lines ($AN$ and $CN$ phase boundaries in Fig.~1a) approach the
same value for $\epsilon \to 0$, i.e. for $P\to P_c$.
This holds for any field direction,
and represents a qualitative difference from the predictions of the $E$
model, where the slope difference approaches a nonzero
limit for in-plane fields
and vanishes for c-axis fields.
 
Our model is further consistent with observed anisotropies of the
penetration depth and $H_{c1}$, and predicts a kink of $H_{c1}$
at $T_*$ for all field directions because of an onset of an additional
order parameter component below $T_*$.

The above given analysis of the phase diagram assumes a homogeneous
symmetry breaking field.
This one-domain picture is an oversimplification of the situation in
UPt$_3$.
Any orthorhombic field is very likely to build a six-fold domain
structure,
as it was observed for the antiferromagnetic order with domain sizes of
less than 150 \AA.
Such an underlying domain structure can
significantly alter the superconducting state \cite{joint}.
This point deserves further investigation to allow a more reliable
comparison with experiment.

Our model may resolve some additional phenomenological puzzles. Most
notable is the absence of Josephson tunneling from conventional
superconductors
into heavy fermion
materials, in particular for UPt$_3$\cite{weaklink}.
Assuming the superconductivity to arise from
FCM pairs with odd-in-${\cal T}$ symmetry, no single pair Josephson
coupling  to a conventional even-in-${\cal T}$ superconductor is
possible without applied or internal (spontaneous) magnetic field that
breaks ${\cal T}$.  
Even in this case the 
staggered phase across an arbitrary sample face will yield  
zero net pair tunneling unless: (i) one cleaves along planes
perpendicular to the pairing $k$-vectors, or (ii) one uses atomic scale
tunneling probes (e.g., STM), which can sample a small region of
non-zero phase.  
This suggests a clear experimental test of the odd-in-${\cal T}$
hypothesis.

We acknowledge useful discussions with M.R. Norman.
This research was supported by a grant from the Department of Energy,
Office of Basic Energy Sciences, Division of Materials Research, and
for R.H. by a grant from the Deutsche Forschungsgemeinschaft.

%----------------------- references --------------------------------

%
%--------------------------- tables --------------------------------
%
\begin{table}
\caption{
Order parameters for stable phases of model (1) for ${\bf H}=0$
depending on the parameter regions I-IV (cf. Fig.~3).
Results apply for a constant (static) strain field $\epsilon$.
Phase 1 appears at $T_+$, phase 2 at $T_*=T_+ -
\beta_1/(\beta_1-\beta_2+\beta_3) (T_+-T_-)$,
$T_*<T_+$, and a possible third phase may exist below
$T_3 =T_+ -2(\beta_1+\beta_2-\beta_3)/(\beta_1-\beta_2-\beta_3)(T_+
 -T_-)$, $T_3 <T_*$.
x and $\phi$ are temperature dependent real numbers.
$\epsilon=0$ corresponds to the high-pressure phase (Fig.~1b).
}
\label{tab1}
\begin{tabular}{c|ccc|c}
region  & \multicolumn{3}{c|}{$\epsilon > 0$} & $\epsilon=0$ \\
 & phase 1 & phase 2 & phase 3 & \\
\tableline
I & (0 0 1) & (0 $i$x 1) & - & (0 $i$ 1) \\
II & (0 0 1) & ($i$x $i$x 1) & ($e^{i\phi}$x $e^{-i\phi}$x 1) &
($e^{i 2 \pi/3}$ $e^{-i 2\pi/3}$ 1 ) \\
III & (0 0 1) & - & - & (0 0 1) \\
IV & (0 0 1) & (x x 1) & - & (1 1 1)

\end{tabular}
\end{table}

%
%----------------------- figure captions ---------------------------
%
%-- fig.1
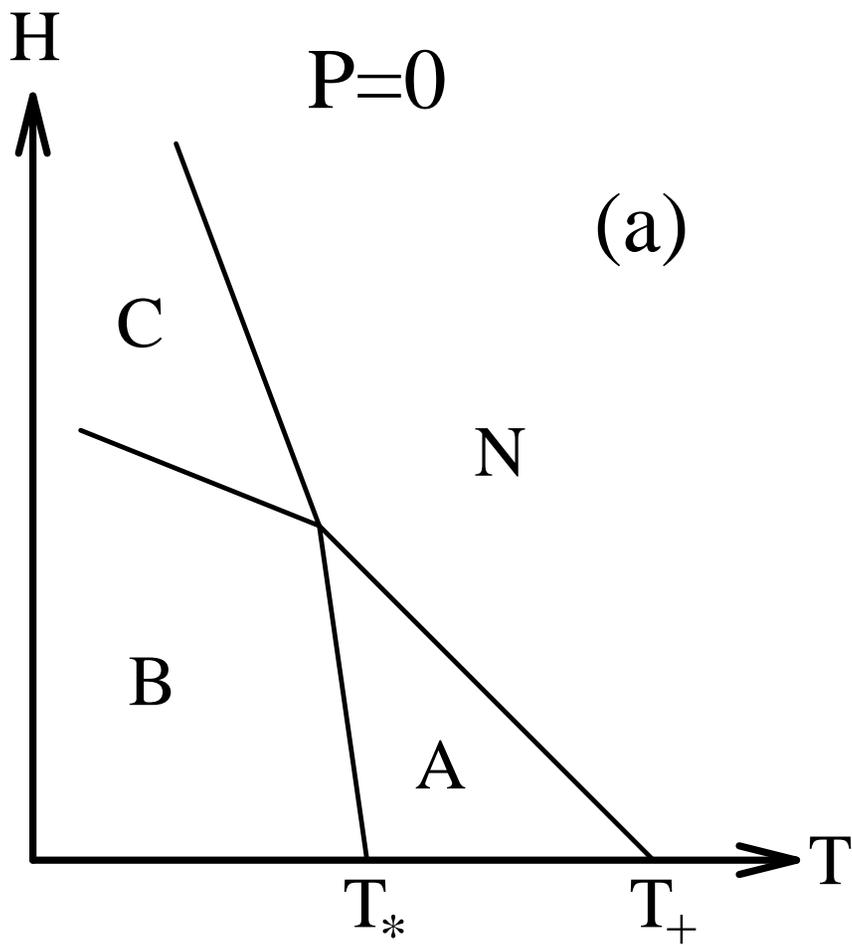
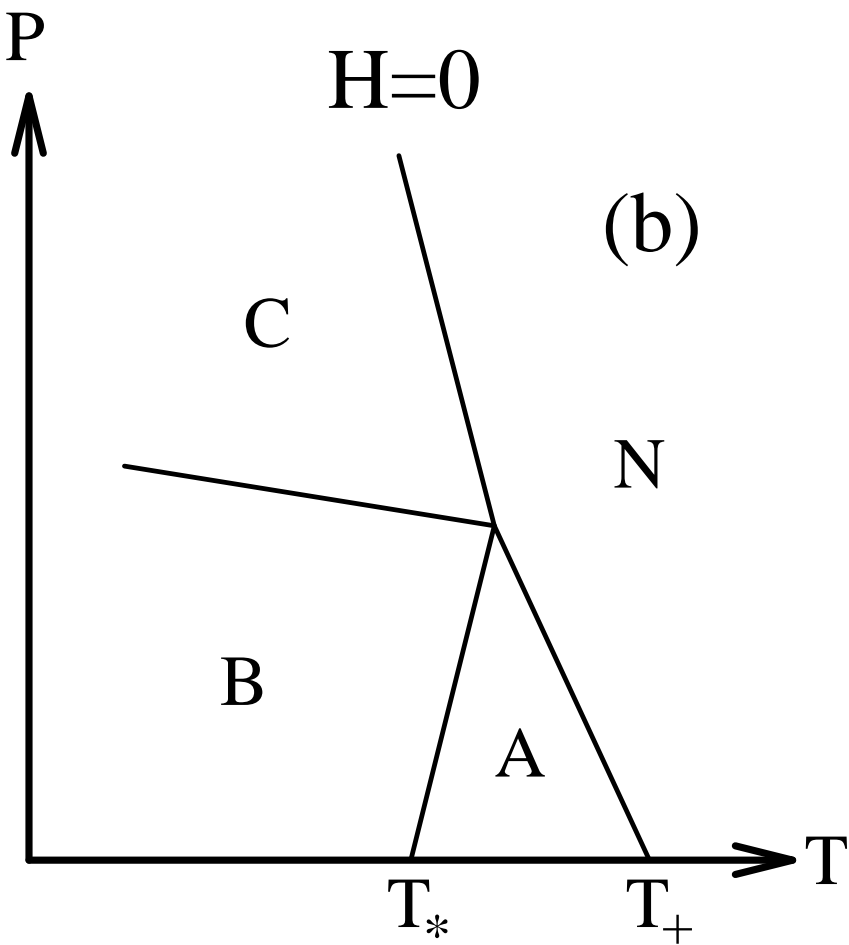
\begin{figure}
\caption{Schematic phase diagram of UPt$_3$ in the $P=0$ and $H=0$
planes as deduced from experiments \protect\cite{adenwella89}.
Areas supposed to belong to the same phase in the three-dimensional
$H$-$P$-$T$ diagram are given the same letter.
}
\end{figure}
%-- fig.2
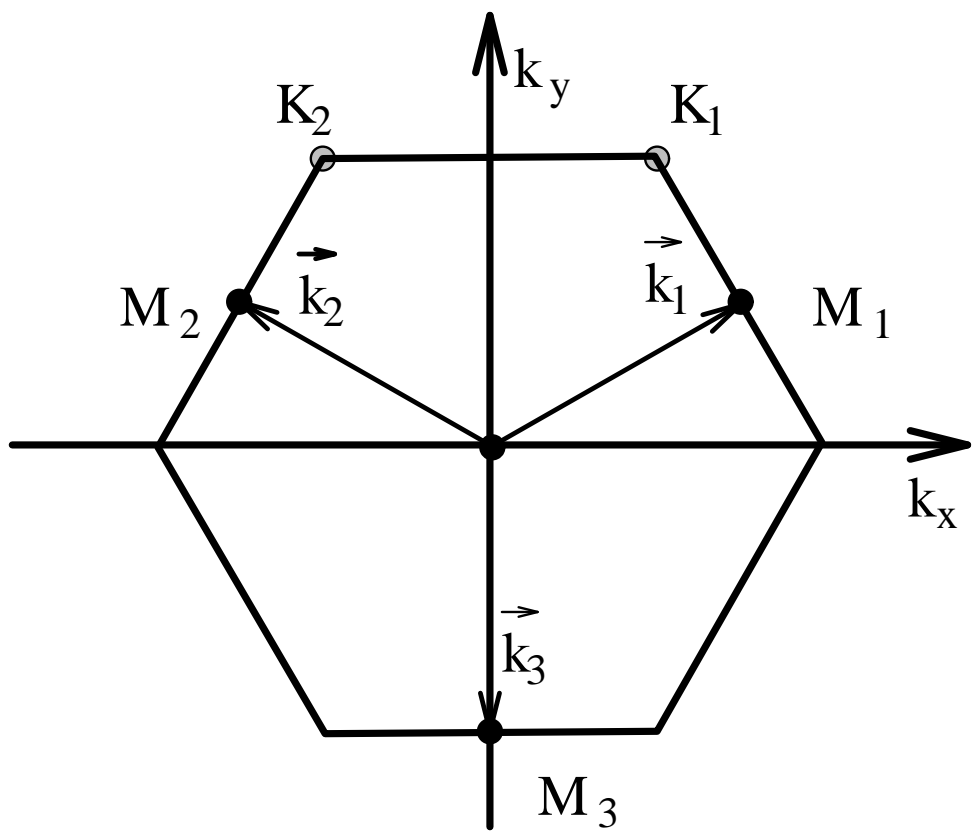
\begin{figure}
\caption{$k_z=0$ cross-section of the Brillouin zone of UPt$_3$.
{$\bf k_i$} are wave vectors of the star of
the M-point representation used in our model.}
\end{figure}
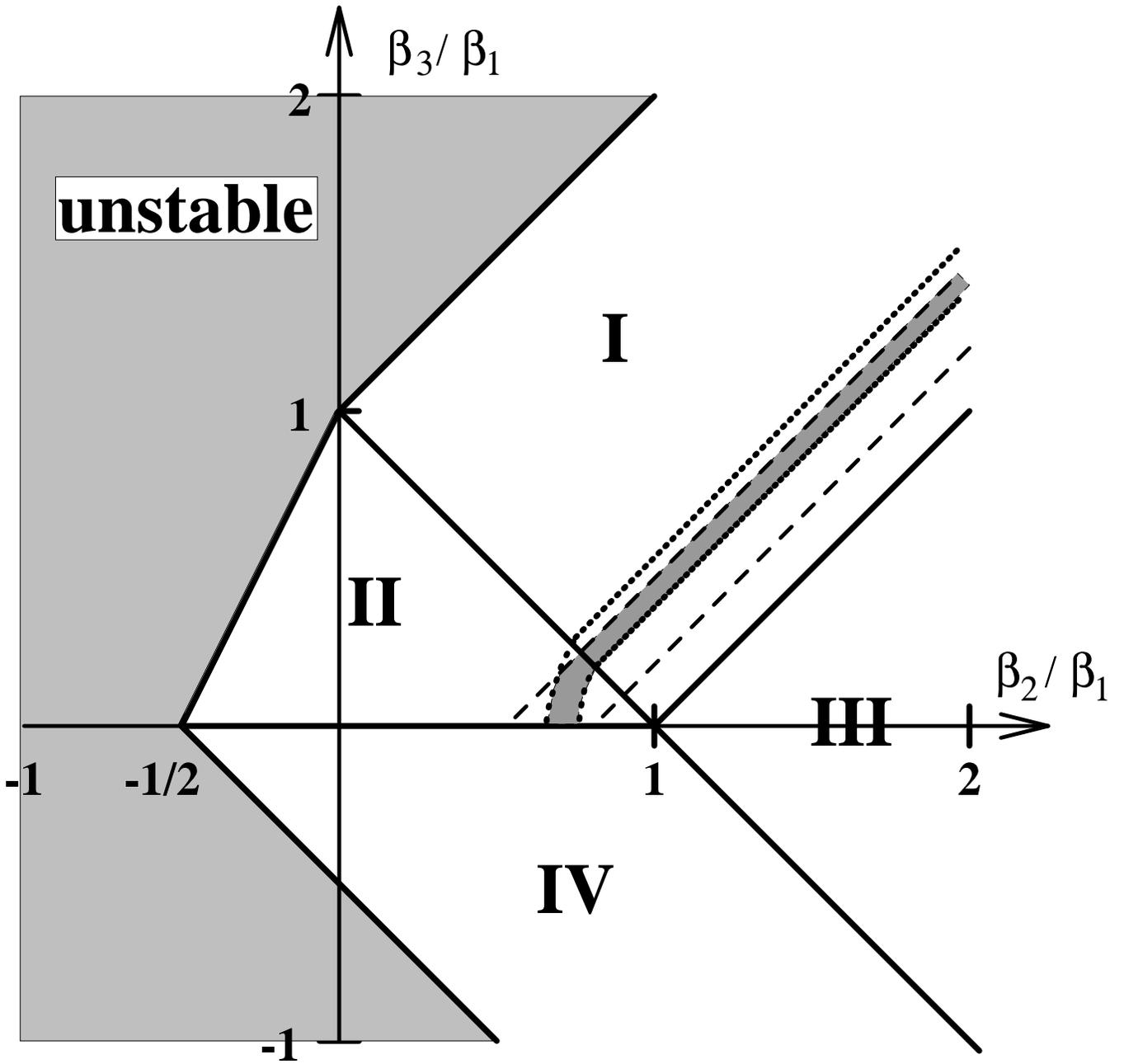
\begin{figure}
%-- fig.3
\caption{Region of stability for solutions of model (1) for $H=0$
as a function of
Ginzburg-Landau parameters for $\beta_1 > 0$.
The free energy has no lower bound in the unstable region, which holds
also for $\beta_1 <0$.
The shaded strip marks the region consistent with specific heat
measurements \protect\cite{fischer89,hasselbach89} (dotted lines: the
specific heat jump ratio, $r$, at $T_*$ and $T_+$ with respect to the
normal state, $1.25 < r < 1.33$),  estimates for $T_-$ from
 $H_{c2}$-measurements \protect\cite{adenwella89} (dashed lines;
$x=(T_+-T_-)/(T_+-T_*), \hspace{2mm} 0.2 < x < 0.45$) and the
assumption that only phase B violates ${\cal T}$-invariance.
Characteristic phase diagrams found in regions (I)-(IV) are described
in the text and in Table 1.}
  \end{figure}

\begin{references}
\bibitem{also} Incommensurate 6- and 12-point star representations are
also possible, but the free energies are actually simpler in form
than that of the $M$-point.
\bibitem{regionlines}
Parameter regions are derived without the more physical condition
$T_*>0$
(or $T_3>0$). Adding this constraint leads to a slight ($T_+$ and
$\epsilon$ dependent) shift of the region boundaries.
\bibitem{bcline}
In this simple picture, the BC line is horizontal which would be at
variance with ultrasound investigations \cite{upt3press}.
However, by replacing the static strain field by a thermodynamic
variable
$M$ describing the AF-magnetic moment ($\epsilon \propto M^2$),
the slope of the BC line can be modified significantly.
\end{references}
\end{document}